# Raman spectra of Nontraditional Compound $KO_4$ and Novel Chemical Reaction of $KCl-O_2$

Yu Tian[a,b,c]*, WanSheng Xiao[a,b], DaYong Tan[a,b], YunHong He[a,b,c], HuiFang Zhao[a,b,c], Feng Jiang[a,b,c]

a CAS Key Laboratory of Mineralogy and Metallogeny, Guangzhou Institute of Geochemistry, Chinese Academy of Sciences, 511 Kehua Street, Guangzhou 510640, China

b Key Lab of Guangdong Province for Mineral Physics and Materials, 511 Kehua Street, Guangzhou 510640, China

c University of Chinese Academy of Sciences, 19 Yuquan Road, Beijing 100049, China

*Corresponding author, e-mail: tianyu@gig.ac.cn

**Abstract:** The non-traditional compound $KO_4$ still existing at ambient conditions was synthesized by the reaction of $KCl-O_2$ under high pressure and high temperature (HPHT) using the diamond anvil cell and the laser heating technology. The $KCl-O_2$ sample was heated at 37 GPa (1800 ± 200 K) and then the products were measured by Raman technology at room temperature. The acquired 1386 $cm^{-1}$ characteristic vibration band of $KO_4$ reflects the O-O pair with fractional negative charge. The other reaction products include another non-traditional compound $KCl_3$, a bit $KClO_4$ and intermediate product $Cl_2$. The $KCl_3$ can distinguish 11 Raman peaks and decompose into KCl and $Cl_2$ below 10 GPa during decompression. These novel products show that high pressure promotes oxygen and chlorine forming unconventional pair-anions and polyanions, the novel chemical reaction provides new perspectives for developing new batteries materials and studying batteries reactions.

## 0 Introduction

Nontraditional oxygen-rich compounds synthesized under HPHT is widely applied to scientific research. For example, $MgO_2$ was used in agricultural and environmental industries for its stably releases oxygen[1]; $FeO_2$ provides a new perspective at oxygen and hydrogen cycling in the deep earth interior[2]; $BeO_2$ also was drawn significant attention, as its band gap becoming larger with increase of pressure [3]. Especially synthesized $LiO_4$ can be treated as electrode material and provides the oxygen source in electrode reaction, which shows great potential in the batteries design and application [4].

In recent years, the metal–air batteries particularly attract attentions owing to the simplicity of the underlying cell reaction. Especially the alkali metal-$O_2$ batteries are the focus of research. Among them, $Li-O_2$ batteries were widely studied because of their small size and high energy storage [5, 6]. $Li-O_2$ system is chosen by more people to study room-temperature batteries [7]. In addition, $Na-O_2$ and $K-O_2$ batteries have also been extensively studied [7-10]. However, electrolyte decomposition is irreversible during cycling in $Li-O_2$ cells with complex chemistry. In recent research, $K-O_2$ cell presents reversible discharging and charging [10], which demonstrates that substitution of lithium by potassium may offer a new perspective in the design of metal-air batteries. Different from gaseous oxygen, high oxygen content solid oxides can provide a stable source of O for batteries reactions, therefore synthesis high oxygen content oxide is imminent in the K-O system, using high temperature and high pressure technology. At present, there are two methods for synthesizing high-oxygen alkali metal oxides under high temperature and pressure, one is obtained by reacting peroxide or superoxide with oxygen($LiO_4$)[4]; the other is obtained by reacting an alkali metal ionic compound with oxygen($NaO_4$)[11]. Because the alkali metal peroxides and superoxide are expensive, difficult to keep and corrosive after absorbing water in the air, less stable than alkali metal ionic compound, the $KO_4$ was synthesized by the chemical reaction of $KCl-O_2$ under HPHT in this work, researched by Raman technology.

# 1 Experiment and calculation

## 1.1 Experimental methods

The high-temperature and high-pressure experimental equipment used in this study was a symmetrical diamond anvil cell (DAC) device, the diamond anvil has a top diameter of 400 μm. A piece of T301 stainless steel foil was pre-indented to a thickness of about 45 μm with 105um diameter size holes drilled in the center of foil as the sample chamber. The KCl sample was pre-pressed to a thickness of about 15 μm using DAC, and select the 50 μm × 50 μm KCl slices were placed in the sample chamber, filled with liquid oxygen joined using liquid nitrogen cooling method. 3~4 ruby chips were put around the KCl sample piece in the sample chamber as the pressure calibration.

The samples were first pressurized to 37.1 GPa at room temperature, then were heated by a double-side SPI fiber laser heating system with 1070 nm wavelength and 100 W power at 1800 ± 200 K about 1 hour, as solid oxygen existing under high pressure absorbing laser energy to generate high temperatures. The radiation spectrum of the sample was collected using a spectrometer and the heating temperature of the sample was fitted by the blackbody radiation equation. After laser heating, the Raman spectra of samples were measured on a Renishaw 2000 Raman spectrometer at room temperature with the 532 nm excitation wavelength. The Raman signal is separated by an 1800 l/mm grating and collected by a thermoelectric cooling CCD.

## 1.2 Calculation methods

The Raman spectra of $KCl_3$ were computed using CASTEP module of Materials Studio software based on density-functional perturbation theory. The Generalized Gradient Approximation (GGA) and Perdew Burke Ernzerhof (PBE) were elected as exchange-correlation functional in this calculation. The Pseudopotential is Norm conserving. Plane wave basis sets with the 830 eV cutoff and a 0.04 Å$^{-1}$ intervals of Monkhorst-Pack scheme for the Brillouin zone are applied in the calculation.

# 2 Results and Discussion

## 2.1 Products of $KCl-O_2$ chemical reaction at HPHT

### 2.1.1 $KO_4$

After laser heating at 37.1 GPa, the pressure of the sample system was reduced to 32.8 GPa. In higher wavenumber area, except the 1621 cm$^{-1}$ (not listed) symmetrical stretching vibration peak of ε-$O_2$ [12], no other peaks were detected. However, as Fig. 1, the sample system was unload to ambient pressure with the gaseous substance being run out, three peaks at 1386 cm$^{-1}$, 1361 cm$^{-1}$ and 1345 cm$^{-1}$ were found on the black products. At ambient conditions, comparing the O-O stretching vibration of $O_2$ at 1556 cm$^{-1}$[13], the O-O stretching vibration of $KO_2$ lies in 1156 cm$^{-1}$[14], the O-O stretching vibration of $K_2O_2$ lies in 762 cm$^{-1}$[15]. The O-O stretching vibration with a negative charge has a longer bonding length than neutral O-O, considered the smaller the amount of charge on oxygen atom in O-O, the greater Raman shift(Fig.2). Therefore, these three Raman peaks may belong to $KO_x$(x>2). Existing research findings show $O_4^-$ anions configurations separated by noble gas at low temperature are equivalent to the negative charge of O-O atoms in the rectangular arrangement ($D_{2h}$), and the infrared spectrum of the O-O stretching vibration of $O_4^-$ anion lies in 1290 cm$^{-1}$ [16], the theoretical calculated O-O stretching vibration of $O_4^-$ anion lies in 1300 cm$^{-1}$ or 1384 cm$^{-1}$ [17]. Combining these two evidences, we can think the observed Raman peak at 1386 cm$^{-1}$ in this study may belong to the O-O stretching vibration of $KO_4$ and the 1361 cm$^{-1}$, 1345 cm$^{-1}$ are the other vibration modes with crystals more complex than single molecules. From this work, the black powder $KO_4$ can be stored at atmospheric pressure. However, it is difficult to detect the Raman bands of $KO_4$ under high pressure for strong interference of deformed diamond Raman peak (1332 cm$^{-1}$).

In Fig. 2, the linear relationship between the Raman shift (just O-O stretching vibration) and amount of charge on oxygen for all the alkali metal peroxides, superoxide and pure oxygen are shown. The Raman wavenumbers [6, 14, 15, 18-20] are listed in table 1 and the Raman shift and O atom charge number maintain good linear consistency(Table 2). We find the straight line slope gradually become bigger with sequence of Na, Li, Rb, K, Cs. The actual Raman shift may not necessarily be on these straight lines, all the Raman shift of $AO_x(x>2)$ must be in the interval from $AO_2$ to $O_2$. There is a certain deviation between the measured Raman and the theory, such as the theoretical Raman shift of $NaO_4$ should be larger than $KO_4$, but the $NaO_4$ Raman(1384 cm$^{-1}$) in our published article be smaller than $KO_4$(1386 cm$^{-1}$).

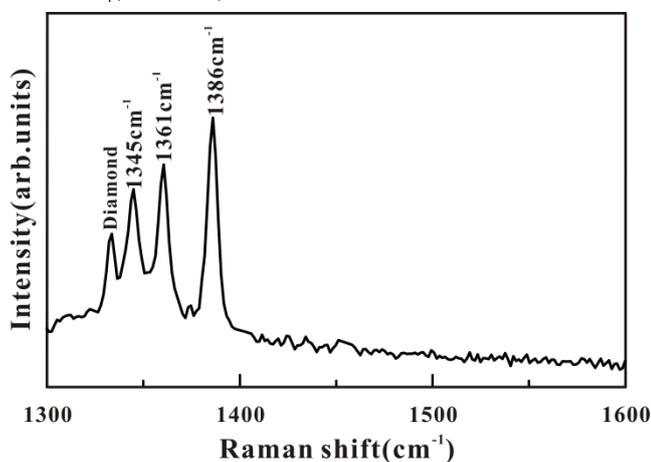

Fig. 1 The characteristic 1386 cm$^{-1}$ Raman peak and other Raman peaks (1361cm$^{-1}$ and 1345 cm$^{-1}$) of $KO_4$ at ambient conditions.

**Table 1 Raman shift(cm$^{-1}$) of stretching O-O vibrations**

| A | Li | Na | K | Rb | Cs |
|---|---|---|---|---|---|
| $A_2O_2$ | 790 [15] | 794 [18] | 762 [15] | 782 [15] | 743 [20] |
| $AO_2$ | 1140 [6] | 1156 [19] | 1142 [14] | 1140 [19] | 1134 [20] |

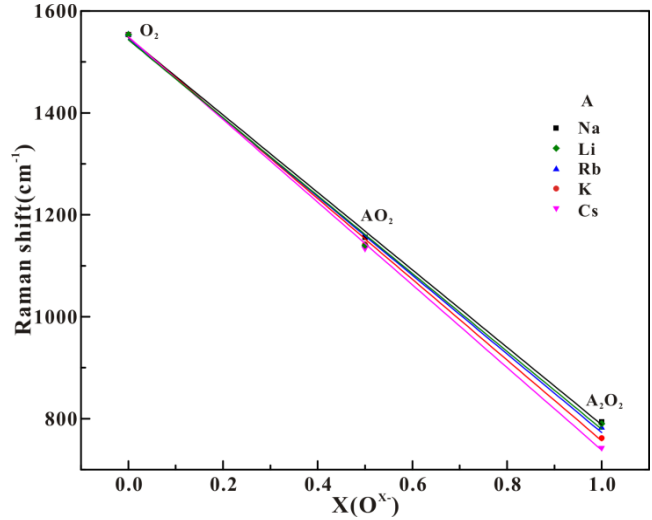

Fig. 2 The relationship between the amount of charge in the O atom in O-O pairs and the Raman shift for alkali metal peroxides, superoxide.

Table 2 The fitting parameters of charge-frequency to a linear equation.[a]

|  | Li | Na | K | Rb | Cs |
|---|---|---|---|---|---|
| a | 1543 | 1548 | 1549 | 1545 | 1549 |
| b | -764 | -760 | -792 | -772 | -811 |
| $R^{2\ b}$ | 0.9977 | 0.9993 | 0.9995 | 0.9982 | 0.9996 |

[a] $\nu = a + b \times X$, $\nu$ is frequency($cm^{-1}$) and X is charge in the O atom; [b] $R^2$ is the fitting coefficient of determination.

### 2.1.2 $KCl_3$, $Cl_2$ and $KClO_4$

As Fig. 3 shows, in lower wavenumber area, the Raman signal of ε-$O_2$ was measured up to 37.1 GPa before heating. The translational vibration peak 225 $cm^{-1}$ ($V_{L1}$) and 490 $cm^{-1}$ ($V_{L2}$) peak, and two Raman peaks (217 $cm^{-1}$ and 615 $cm^{-1}$) that appear only at higher pressures are shown with 1628 $cm^{-1}$ symmetrical stretching vibration peak not listed [12]. In addition, no other Raman signals were detected, reflecting no chemical reaction of KCl and $O_2$ at normal temperature and high pressure. Under this pressure, KCl is a cubic phase with no Raman signal. As the curve a shows, except a series of new Raman peaks in the low-wave region and the 475 $cm^{-1}$ of ε-$O_2$, a 1066 $cm^{-1}$ weak peak presents in the Raman chart. This 1066 $cm^{-1}$ peak was saved to 0.1 MPa after decompressing, as $\nu_1$ mode showing in the Fig. 4. The $\nu_1$ and other peaks ($\nu_2$, $\nu_3$, $\nu_4$) are in good agreement with the results of $KClO_4$ chemical reagents and literature [21], which indicates $KClO_4$ existing in chemical reaction products. When loaded to 18.0 GPa, as the curve b, a 541 $cm^{-1}$ peak shown in the chart except these new Raman peaks found in curve a. The 541 $cm^{-1}$ peak can also be continued to lower pressure (6.8 GPa), as $A_g$-1-2 modes showing in the Fig. 4. The $A_g$-1-2 modes and other peaks($A_g$, $B_{3g}$) are in good agreement with the Raman spectra of pure solid $Cl_2$(space group *Cmcm*) samples in literature[22], which indicates the $Cl_2$ existing in the chemical reaction products. The Raman spectra of $Cl_2$ were collected in large amounts of $Cl_2$ aggregates with high intensity. The red curve c at 22 GPa is the Raman signal of *P-3cl*-$KCl_3$ referred from Zhang et al.[23], which is consistent with curve b in our experiment, the *P-3cl*-$KCl_3$ may be also the chemical reaction products.

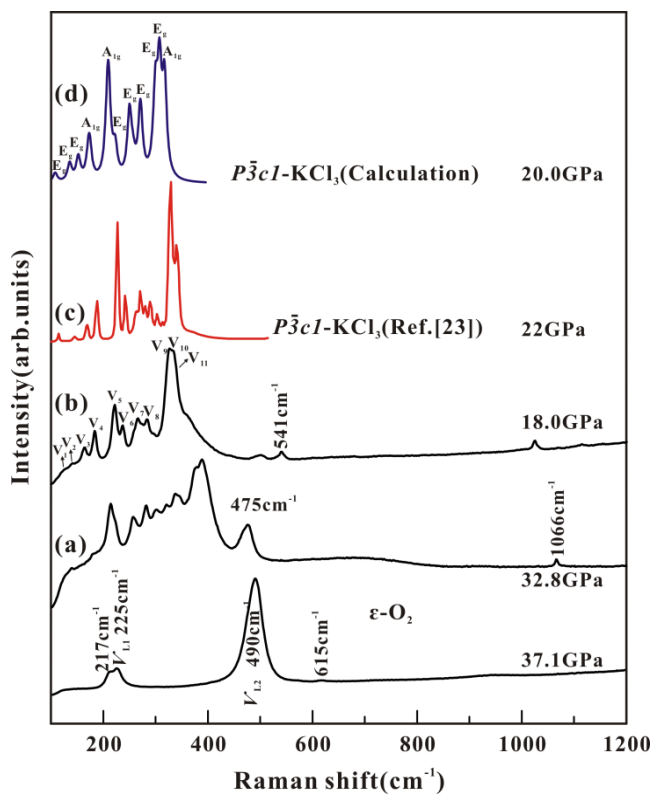

Fig. 3 Compare the Raman spectra (a, b) respectively measured at 37.1 GPa and 18.0 GPa in this experiment, the Raman spectrum (d) calculated at 20.0 GPa in this work with the Raman spectrum (c) of (*P-3c1*) KCl₃ observed at 22 GPa in Ref.[23]

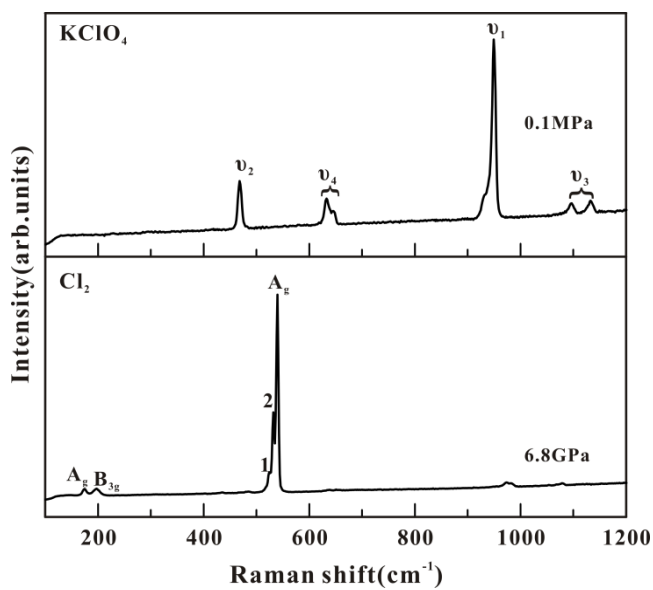

Fig. 4 Raman spectra of *Cmcm*-Cl$_2$ under 6.8 GPa and KClO$_4$ under 0.1 MPa during decompression.

According to group theory, *P-3cl*-KCl$_3$ has 16 Raman active modes ($\Gamma = 5A_{1g} + 11E_g$). However, just eleven Raman peaks are detected in this experiment for *P-3cl*-KCl$_3$, marked as Vi (I = 1, 2, 3….). The First-principles calculated Raman spectroscopy of *P-3cl*-KCl$_3$ shows in curve d of Fig. 3 at 20 GPa, which Raman bands are consistent with experimental result with some reasonable pressure deviation (Table 3). Based on theoretical calculation, the Raman peaks of *P-3cl*-KCl$_3$ can be assigned in table 3. From the structural analysis, the non-traditional compounds *P-3cl*-KCl$_3$ has a non-linear, symmetric Cl-Cl-Cl configuration(Cl$_3^-$) with a negative charge non-uniformly distributed among these three Cl atoms[23], which vibrations is the *P-3cl*-KCl$_3$'s Raman origin. The synthesis of *P-3cl*-KCl$_3$ means the trichloride polyanions, being more difficult to form under atmospheric pressure [24], are easily synthesized under high pressure. The *P-3cl*-KCl$_3$ gradually weakened and disappeared below 10 GPa, as decomposing into KCl and Cl$_2$, which is consistent with Zhang et al. [23]'s experimental results. Therefore, more chlorine was detected at low pressure (6.8 GPa).

Table 3 Raman shift(cm$^{-1}$) of KCl$_3$ between experiment(18.0 GPa) and calculations (20.0 GPa)

|       | V$_1$ | V$_2$ | V$_3$ | V$_4$ | V$_5$ | V$_6$ | V$_7$ | V$_8$ | V$_9$ | V$_{10}$ | V$_{11}$ |
|-------|-------|-------|-------|-------|-------|-------|-------|-------|-------|----------|----------|
| Modes | $E_g$ | $E_g$ | $E_g$ | $A_{1g}$ | $A_{1g}$ | $E_g$ | $E_g$ | $E_g$ | $E_g$ | $E_g$ | $A_{1g}$ |
| Calc. | 108 | 136 | 152 | 173 | 209 | 221 | 250 | 271 | 301 | 307 | 316 |
| Exp.  | 125 | 142 | 164 | 183 | 221 | 237 | 266 | 283 | 326 | 332 | 338 |

## 2.2 Chemical reaction equation of KCl-O$_2$ at HPHT

From the above Raman measurement, we believe that the KCl-O$_2$ system undergoes a chemical reaction under the experimental temperature and pressure. The possibly existing high pressure chemical reactions list following:

$$2KCl + 4O_2 \rightarrow 2KO_4 + Cl_2 \quad (1)$$

$$3KCl + 4O_2 \rightarrow KCl_3 + 2KO_4 \quad (2)$$

$$KCl + 2O_2 \rightarrow KClO_4 \quad (3)$$

Reaction (1) is likely to be an intermediate reaction, resulting in a further chemical reaction of Cl$_2$ with KCl:

$$KCl + Cl_2 \rightarrow KCl_3 \quad (4)$$

Chemical reaction in decompression is shown as:

$$KCl_3 \rightarrow KCl + Cl_2 \quad (5)$$

The chemical reaction (2) is the conjugation reaction, also the confederation of reaction (1) and reaction (4). In KCl$_3$ and KO$_4$, both O and Cl have negative fractional charges. The high pressure favors the presence of O-O pair-anions and Cl-Cl-Cl polyanions with the unconventional chemical valence states of O and Cl.

The reaction (3) is the minor reaction with the lower Raman intensity of less KClO$_4$. The studies of Walker et al. [25] shows that KClO$_4$ decomposes to KCl and O$_2$ at high temperatures of 1.5-9 GPa, which is opposite to the results of high pressure experiments in this study. From another point of view, the formulas (1) and (2) indicate that Cl$^-$ loses electrons to zero or positive valence, while O atoms obtains electrons to negative valence, reflecting O has stronger electronic acquisition capability than Cl under certain pressure conditions.

According to all the chemical reactions, producing the KO$_4$ in KCl-O$_2$ system is rational and believable. However, the structure and properties of KO$_4$ is unknown. In recent research, Yang et al. [4] thinks LiO$_4$ consist of a Li layer sandwiched by an O ring structure inheriting from high pressure ε-O$_2$. Theoretical research shows that LiO$_4$ can be also stable under normal pressure and has superconductivity [26]. Whether there are similar structures and properties on KO$_4$ with LiO$_4$ needs further investigation.

## 3  Conclusions

In summary, we synthesized KO$_4$ still existing at ambient conditions, and investigated the novel chemical reaction of KCl-O$_2$ system under HPHT, which provides new perspectives for developing new batteries materials and studying batteries reactions. In the reaction products, the oxygen and chlorine elements have a negative charge of the fraction, indicating that the two elements have unconventional chemical valence under high pressure, and its mechanism is that high pressure prompts the presence of O-O pair-anions and Cl-Cl-Cl polyanions.

**Acknowledgements**

This work was supported by the Strategic Priority Research Program (B) of the Chinese Academy of Sciences (XDB18010403), and the National Natural Science Foundation of China (41572030).